\documentclass[%
 reprint,
superscriptaddress,
nofootinbib,
 amsmath,amssymb,
 aps,
]{revtex4-2}

\usepackage[dvipsnames]{xcolor}
\definecolor{betterblue}{RGB}{7, 56, 176} 

\setcitestyle{authoryear,round}
\usepackage[colorlinks=true]{hyperref}
\hypersetup{
    urlcolor=black,
    citecolor=betterblue,
    linkcolor=betterblue,
  }

\usepackage{graphicx}
\usepackage{dcolumn}
\usepackage{bm}
\usepackage[mathlines]{lineno}

\usepackage[normalem]{ulem}
\usepackage{subfigure}

\newcommand{\LCDM}{$\Lambda$CDM}
\newcommand{\aap}{A\&A}
\newcommand{\apjs}{ApJS}

\newcommand{\mnras}{MNRAS}
\newcommand{\physrep}{Phys. Rep.}

\newcommand{\apjl}{ApJ}
\newcommand{\pasj}{PASJ}

\begin{document}

\title{Role of the Hubble Scale in the Weak Lensing \textit{vs.} CMB Tension}

\author{Lucas F. Secco}\email{secco@uchicago.edu}
\affiliation{Kavli Institute for Cosmological Physics, Enrico Fermi Institute, The University of Chicago, Chicago, IL 60637, USA}

\author{Tanvi Karwal}
\affiliation{Center for Particle Cosmology, Department of Physics and Astronomy,
University of Pennsylvania, Philadelphia, PA 19104, USA}

\author{Wayne Hu}
\affiliation{Kavli Institute for Cosmological Physics, Enrico Fermi Institute, The University of Chicago, Chicago, IL 60637, USA}

\affiliation{Department of Astronomy \& Astrophysics, The University of Chicago, Chicago, IL 60637,
USA}
\author{Elisabeth Krause}
\affiliation{Department of Astronomy/Steward Observatory, University of Arizona,
933 North Cherry Avenue, Tucson, AZ 85721-0065, USA}

\date{\today}

\begin{abstract}
We explore a re-parameterization of the lensing amplitude tension between weak lensing (WL) and cosmic microwave background (CMB) data and its implications for a joint resolution with the Hubble tension.
Specifically, we  focus on 
the lensing amplitude 
over a scale of 12 Mpc in absolute distance units
using a derived parameter $S_{12}$
and show its constraints from recent surveys in comparison with Planck 2018. 
In WL alone, we find that the absolute distance convention correlates 
$S_{12}$ with $H_0$. 
Accounting for this correlation in the 3D space $S_{12}\times \omega_m \times h$ reproduces the usual levels of $2\sim 3\sigma$ tension inferred from $S_8\times\Omega_m$.
Additionally, we derive scaling relations in the $S_8\times h$ and $S_{12}\times h$ planes that are allowed by $\Lambda$CDM and extrapolate target scalings needed to solve the $H_0$ and lensing-amplitude tensions jointly in a hypothetical beyond-$\Lambda$CDM model. 
As a test example, we quantify how the early dark energy scenario compares with these target scalings.  
Useful fitting formulae for $S_8$ and $S_{12}$ as a function of other cosmological parameters in $\Lambda$CDM are provided, with 1\% precision.

\end{abstract}

\maketitle

\section{Introduction}

Current constraints on cosmological parameters from different experiments point to the possible inadequacy of $\Lambda$CDM as the complete description of the Universe, assuming experimental and astrophysical systematics are negligible. 
Several advances have been made in addressing these tensions on multiple fronts: 
theoretical developments in both modelling data and modifying \LCDM, the acquisition of more and better data, and considerations of the statistical properties of parameter mismatches. 

Within the context of the latter approach, \cite{Sanchez2020} proposed a re-parameterization of the commonly used lensing amplitude parameter $S_8$ (defined below), 
arguing for a change of units in the $\sigma_8$ parameter which is defined as the RMS fluctuation of the linear density field within a spherical top-hat of radius $8$ Mpc/$h$. 
Specifically, their proposal adopts Mpc units  instead of the common Mpc/$h$ choice, with $h$ being defined via the Hubble constant $H_0\equiv 100h$ km/s/Mpc.  
At a reference value of $h\approx 0.67$, the usual $(8/h)$ Mpc scale of fluctuations becomes approximately 12 Mpc. 
This leads to a convenient parameter definition $\sigma_{12}$, analogous to $\sigma_8$ but over a top-hat of radius 12 Mpc without the typical $1/h$ factor, and its corresponding lensing amplitude parameter $S_{12}$. This parametrization has been used, for instance, in deriving cosmological parameter constraints from galaxy clustering and complementary probes as shown in \citet{Semenaite21} and references therein. 
In what follows, we refer to the Mpc/$h$ convention as \textit{relative} distances (i.e.\ relative to the Hubble length or extragalactic distance scale) and the Mpc convention as \textit{absolute} distances.

Part of the motivation for this definition is that, because posteriors of different astrophysical probes might or might not constrain $h$, the parameter $\sigma_8$ is not guaranteed to reference the same absolute scales across experiments. 
Therefore, the potential benefit of the $\sigma_{12}$ convention is that the RMS power spectrum is measured at a fixed, known scale of 12 Mpc regardless of the particular constraint on $h$ of any probe.  A drawback is that low redshift experiments best measure relative quantities through the conversion of redshifts and angles using the distance scale and hence 12 Mpc may not correspond to the best measured aspect of a given survey.

In this work, we investigate whether the proposed change in units and the behavior of the newly  introduced parameters $\sigma_{12}$ and $S_{12}$ impact the degree of agreement/disagreement between a subset of recent weak-lensing 2-point correlation results (also commonly referred to as cosmic shear) (\cite{Hikage2019,Hamana2020,Asgari2021,Amon2022}, \citet*{Secco2022}), and Planck 2018 \citep{Planck2018parameters}, and secondly, whether these new parameters aid physical interpretation. 
As we show, transitioning from $S_8$ to $S_{12}$ naturally introduces the Hubble parameter, so we additionally explore their functional dependence and relations of the form $S_{8}(h,\ldots)$  and $S_{12}(h, \ldots)$
to gain insight into solutions of the apparent tensions in both the lensing amplitude and Hubble parameter. 

This paper is organized as follows: in Section II, we analyze how tension metrics change under $S_{12}$; in Section III we interpret those tension metrics and obtain fitting formulas for the scaling of lensing amplitude parameters with $h$ in $\Lambda$CDM; in Section IV we explore the desirable scalings of non-$\Lambda$CDM models that could potentially solve tensions in both parameters jointly, and examine the particular case of a model with an early dark energy (EDE) component behaving as a cosmological constant at early redshifts and transiently affecting the expansion rate around matter-radiation equality \citep{Karwal2016,Poulin2019,Lin2019,Smith2020,Poulin:2018dzj,Agrawal:2019lmo}, finally concluding in Section V.

\section{Cosmic Shear Constraints}
\label{sec: postprocessing}

\begin{figure*}
	\includegraphics[width=\columnwidth]{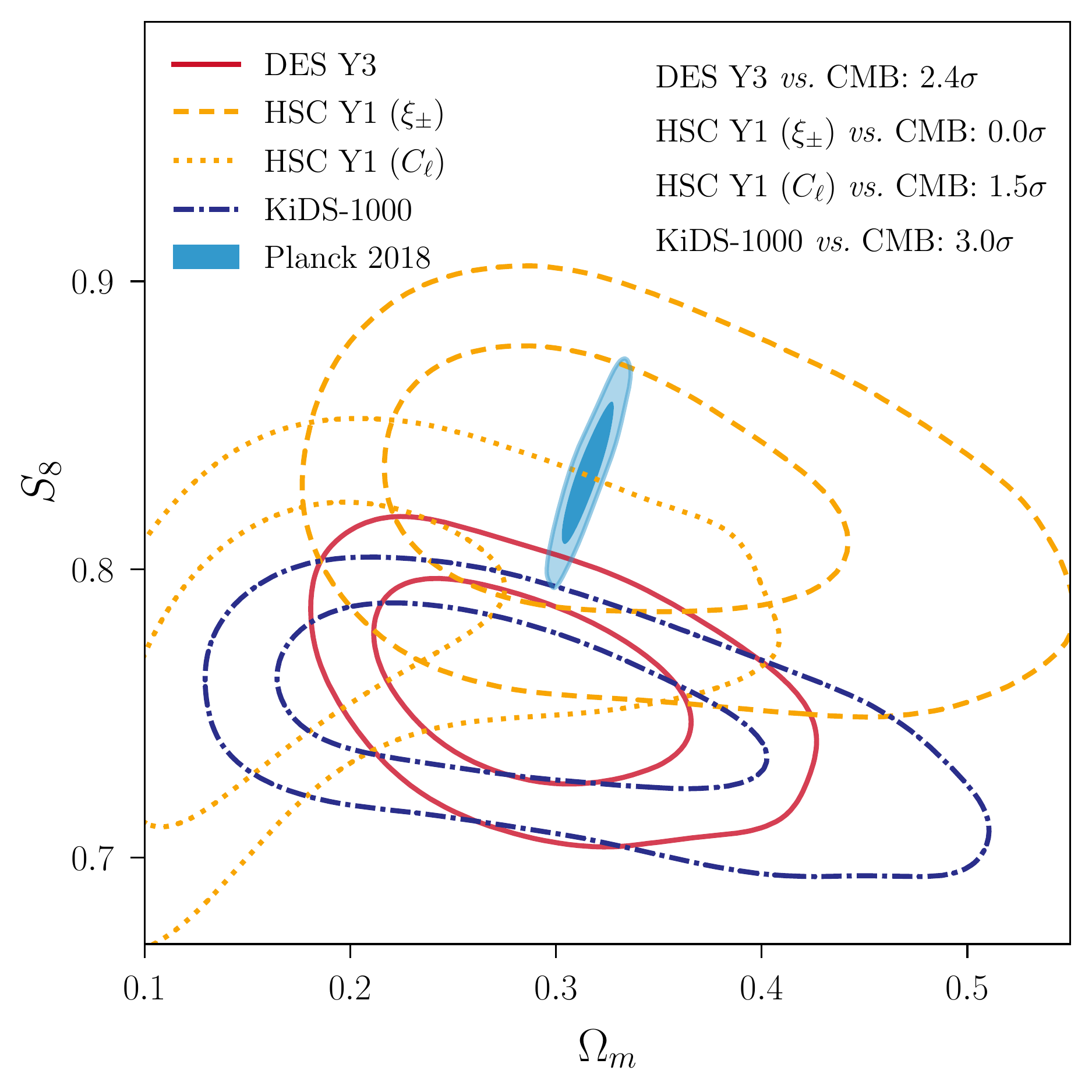}
    \includegraphics[width=\columnwidth]{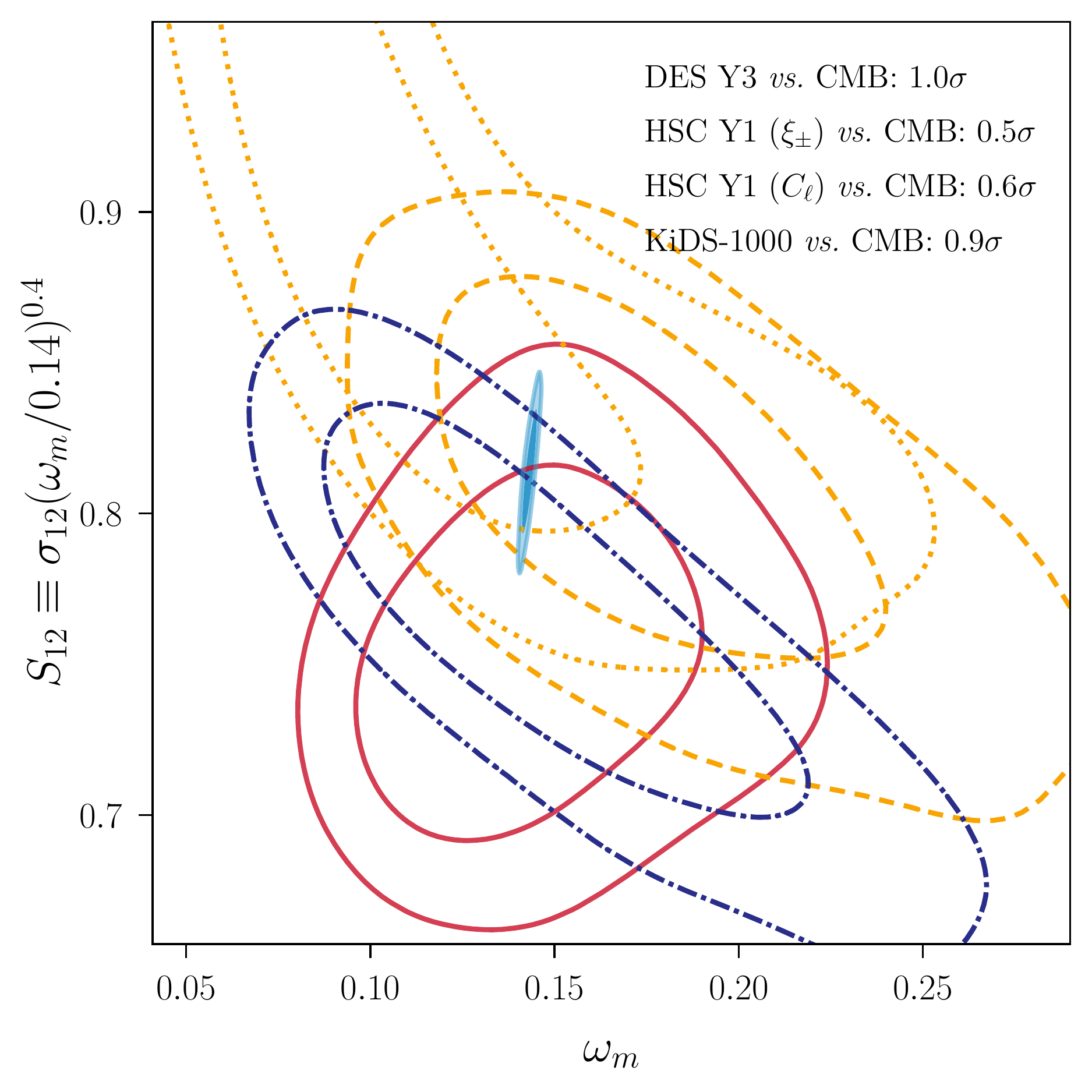}
    \caption{
    \textbf{Left panel:} 
    Nominal constraints in the $S_{8}\times\Omega_\mathrm{m}$ subspace as inferred from different weak-lensing surveys (DES Y3, HSC Y1 and KiDS-1000) compared with Planck 2018. 
    Numbers on the top right correspond to the $Q_{\mathrm{DM}}$ metric estimated for each survey in this 2D parameter subspace. 
    \textbf{Right panel:} 
    Same as left panel, but in the newly-derived $S_{12}\times\omega_\mathrm{m}$ plane. 
    In comparison with the $S_8\times\Omega_m$ results of each survey, we note that the discrepancy between shear surveys and the CMB appears to be significantly reduced. Note that the marginal $S_{12}$ constraint, unlike $S_8$, depends strongly on each survey's $H_0$ prior (see Sec. \ref{sec: full parameter space}), complicating consistency statements based on the $S_{12}$ posterior alone.}
    \label{fig: baseline}
\end{figure*}

In order to study the lensing amplitude tension under the newly-defined $S_{12}$, 
we first derive it from the publicly-available Markov Chain Monte Carlo (MCMC) results where it was not a standard output, namely in 
KiDS-1000 (COSEBIs)\footnote{\url{https://kids.strw.leidenuniv.nl/DR4/KiDS-1000_cosmicshear.php}} \citep{Asgari2021} and 
HSC-Y1\footnote{\url{http://th.nao.ac.jp/MEMBER/hamanatk/HSC16aCSTPCFbugfix/index.html}}\footnote{\url{http://gfarm.ipmu.jp/~surhud/PDR1_HSCWL/Hikage/}} \citep{Hikage2019,Hamana2020}. 
We employ \textsc{CosmoSIS}\footnote{\url{https://cosmosis.readthedocs.io/en/latest/}} \citep{zuntz15} to read each existing sample in the public chains as an input cosmology, run the Boltzmann solver \textsc{CAMB}\footnote{\url{https://camb.info/}} \citep{camb} and compute
\begin{equation}\label{eq: sigma12}
    \sigma_{12}^{2}\equiv\int\frac{d^{3}k}{(2\pi)^{3}}\left|W_{12\,\mathrm{Mpc}}(k)\right|^{2}P_{\mathrm{lin}}(k) \,,
\end{equation}
where $W_{12}(k)$ is the Fourier transform of a top-hat filter with radius 12 Mpc and $P_{\mathrm{lin}}$ is the linear matter power spectrum. 
The definition above is analogous to that of the usual $\sigma_8$ parameter which employs distance units relative to $h$ (thus with wavenumbers $k$ also carrying a factor of $h$):
\begin{equation}
        \sigma_{8}^{2}\equiv\int\frac{d^{3}k}{(2\pi)^{3}}\left|W_{8/h\,\mathrm{Mpc}}(k)\right|^{2}P_{\mathrm{lin}}(k) \,.
\end{equation}
We additionally derive the corresponding lensing amplitude parameter $S_{12}$ as proposed in \cite{Sanchez2020}, in contrast with the usual definition of $S_8$:
\begin{equation}\label{eq: S12 definition}
    S_{12}\equiv \sigma_{12}\left(\frac{\omega_m}{0.14}\right)^{0.4} \,,
\end{equation}
\begin{equation}\label{eq: S8 definition}
    S_{8}\equiv \sigma_{8}\left(\frac{\Omega_m}{0.3}\right)^{0.5} \,,
\end{equation}
with $\omega_X=\Omega_X h^2$ defined as the physical density parameter at redshift zero for a component $X$.
All original parameters and sampling weights are left unchanged. 
The same postprocessing is not required for DES Y3 
(\cite{Amon2022}, \citet*{Secco2022})
and Planck 2018 temperature and polarization results \citep{Planck2018parameters}, as those public chains\footnote{Chains for DES Y3 and Planck 2018 TTTEEE, low EE and low TT including $\sigma_{12}$ are taken from \url{https://des.ncsa.illinois.edu/releases/y3a2/Y3key-products}.
We further postprocess the Planck chain to approximately fix the neutrino mass to minimum. } 
already contain the extra derived parameter in Eq.~\eqref{eq: sigma12}. 

To compare surveys with the CMB, we compute a $Q_\mathrm{DM}$ Gaussian tension metric based on shifts between parameter posteriors \citep{Raveri2020}, carried out with the publicly available \textsc{tensiometer} code\footnote{\url{https://github.com/mraveri/tensiometer}} \citep{Raveri2019,Raveri_Doux_2021}. We initially inspect that metric in the 2D subspaces that are best constrained in weak lensing, namely $S_{12}\times\omega_m$ and $S_8\times \Omega_m$.
This choice is a starting point for the present analysis based on the experience that the $S_8\times \Omega_m$ plane contains most of the tension information when comparing weak lensing to the CMB. It will become clear that the same cannot be said about the $S_{12}\times \omega_m$ subspace, so we shall look more broadly at the full parameter space later on, in Sec. \ref{sec: full parameter space}.

With $S_{12}$ derived for all data sets, we compare
the lensing amplitude constraints in the 
$S_{12}\times\omega_m$ plane  with the standard  $S_8\times\Omega_\mathrm{m}$ plane in Fig.\ \ref{fig: baseline}.
No re-analyses or modifications to priors and nuisance parameters are implied.  
Numbers on the top right correspond to the Gaussian $Q_\mathrm{DM}$. 
As each published analysis quantifies tension with different metrics, we do not expect to exactly reproduce their reported numbers, but simply seek to reasonably approximate them.

\begin{figure*}
    \includegraphics[width=\columnwidth]{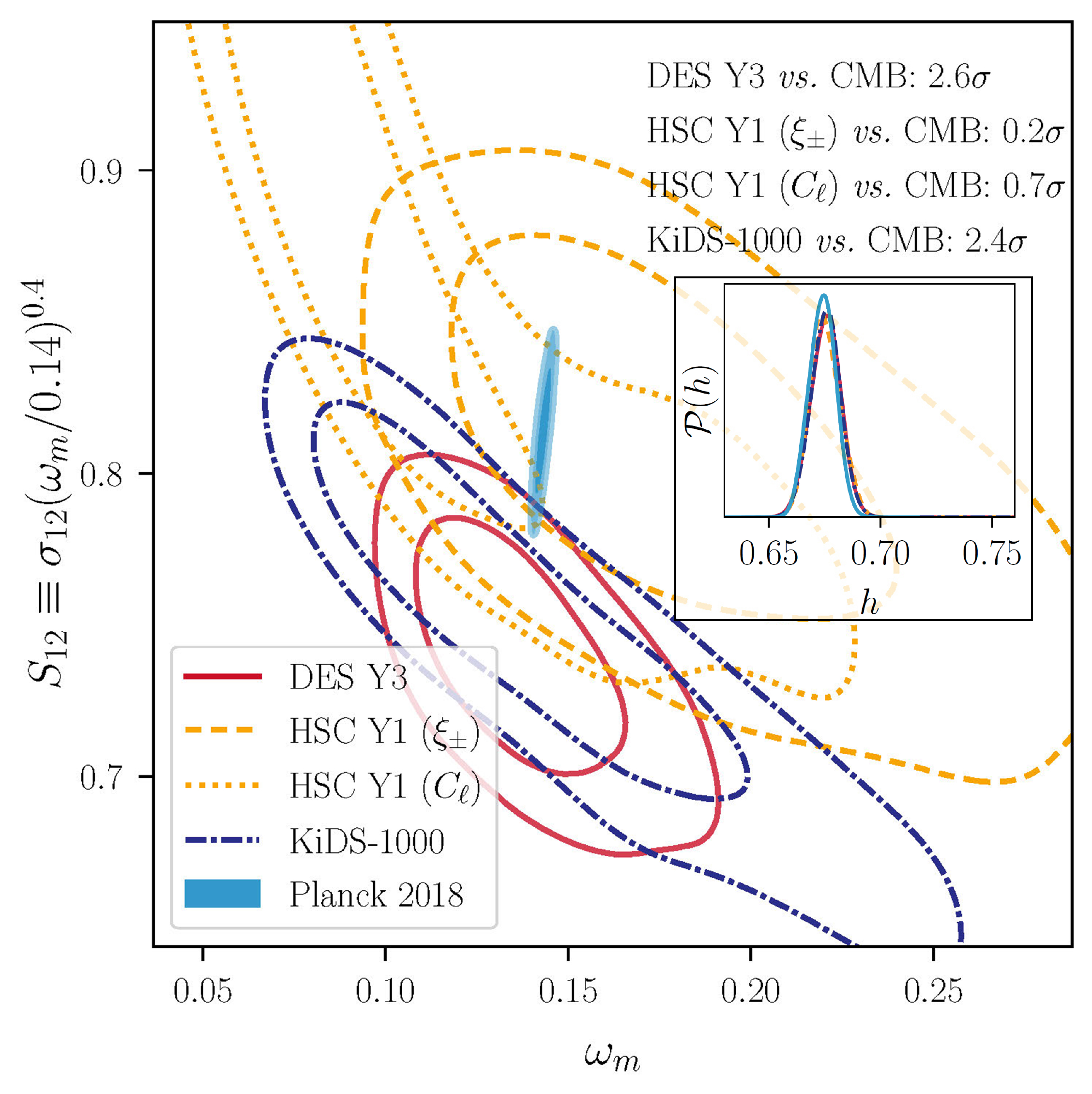}
    \includegraphics[width=\columnwidth]{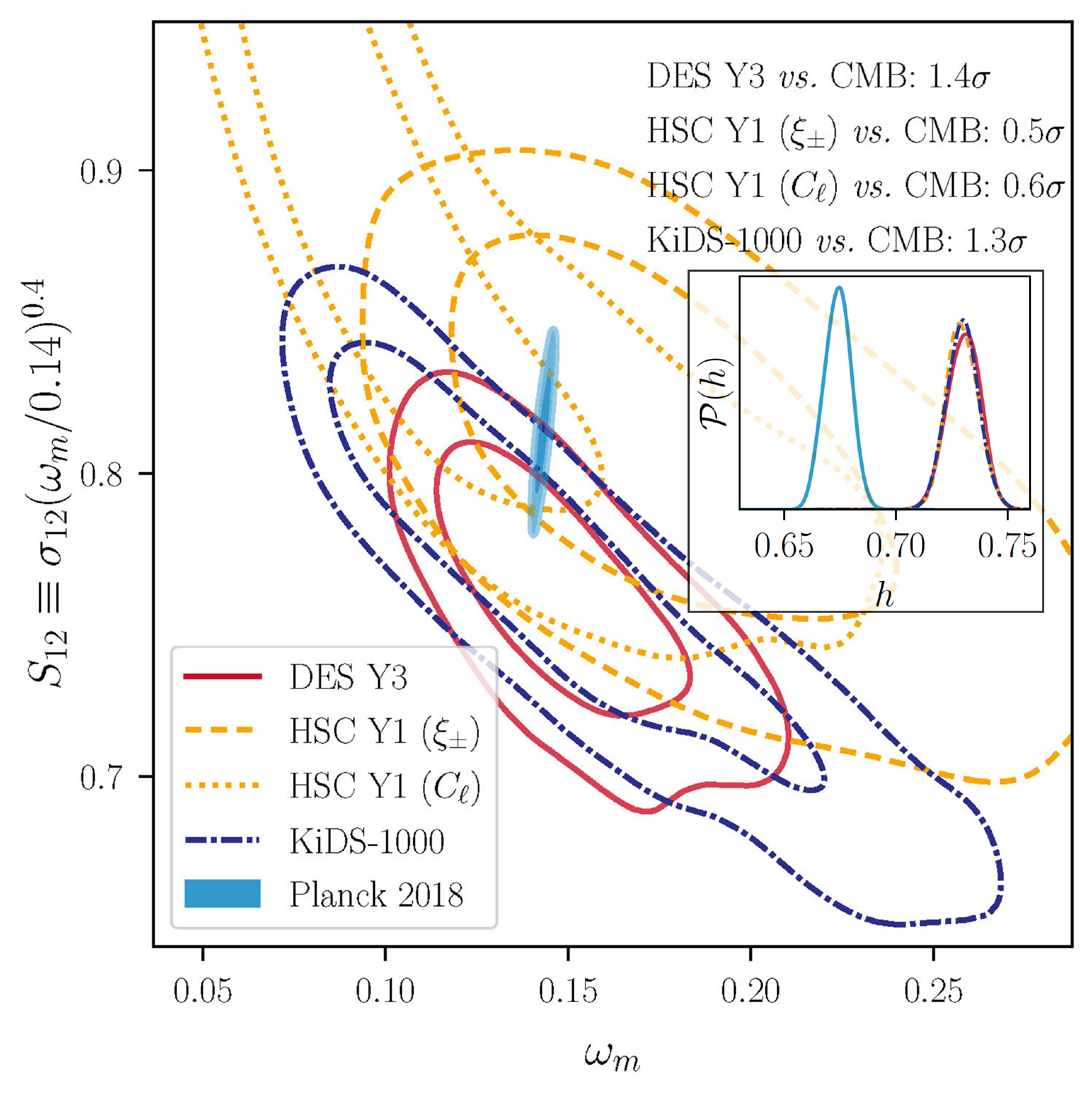}
    \caption{Constraints on the $S_{12}\times\omega_m$ plane from different weak-lensing surveys after introducing informative priors on $h$. 
    Numbers in the upper-right corner are $Q_{\mathrm{DM}}$ statistics in the subspace being plotted. 
    Note that the Planck contour is not re-sampled in either case.  
    \textbf{Left panel:} 
    A Hubble parameter prior centered at $h=0.673$, in agreement with Planck 2018 and with comparable constraining power. 
    We find that the $Q_{\mathrm{DM}}$ tension approaches the values obtained in the $S_8\times\Omega_m$ plane by each surveys' reported analyses. 
    \textbf{Right panel:} 
    Hubble parameter prior centered at $h=0.73$. 
    Tension metrics remain lower than each survey's nominal result (see left panel of Fig.\ \ref{fig: baseline}). 
    Note that this prior introduces a much larger tension in the full parameter space (see Sec. \ref{sec: full parameter space}).
    }
    \label{fig: postprocessed}
\end{figure*}

The most remarkable aspect of the $S_{12}\times\omega_m$ panel of Fig.\ \ref{fig: baseline} is the apparent absence of a statistically significant tension between CMB and lensing data sets. 
As we shall quantify in Section \ref{sec: LCDM scalings}, the choice of absolute Mpc units creates a dependence between $S_{12}$ and $h$ when well-measured lensing parameters are held fixed. 
Thus marginalization over the wide, but still informative $h$ prior\footnote{For reference, KiDS and the HSC real-space analysis adopt prior bounds $h=[0.64, 0.82]$, the harmonic-space analysis of HSC adopts $h=[0.60, 0.90]$ and DES adopts $h=[0.55, 0.91]$, all of which are uniformly distributed.} adopted by lensing surveys not only dilutes the cosmic shear constraining power in the $S_{12}\times\omega_m$ subspace but also impacts the $S_{12}$ posterior. It will therefore become clear that it is impossible to assess consistency between surveys using $S_{12}$ without matching their Hubble priors. 

It is important to note that $h$ is largely prior-dominated in cosmic shear probes alone, and therefore the usual $S_8$ parameter in relative distance units \citep{Jain_Seljak_1997} is conveniently constructed to \textit{not} be similarly sensitive to constraints on the Hubble parameter. 
More recently, using a halo-model approach, \cite{Hall2021} showed useful parameter scalings for cosmic shear that clarify this probe's insensitivity to $H_0$, pointing out that external data (or possibly higher order lensing correlations) are needed to significantly break parameter degeneracies and lead to a lensing-based measurement of $H_0$ with precision comparable to the current state-of-the-art.

Since $S_{12}$ constraints are diluted due to the marginalization over the Hubble parameter and dependent on its prior boundaries, could the addition of an external informative prior on $h$ mitigate those effects?
We explore this consideration below, with two alternatives. 

\subsection{An External $H_0$ Prior in Agreement with Planck 2018}\label{sec: planck prior}

We impose an informative prior on the Hubble parameter in the existing DES Y3, HSC-Y1 and KiDS-1000 chains via importance sampling. 
Specifically, we modify the weights of MCMC samples following a desired distribution, which we take to be a Gaussian centered at $H_0=67.3$ km/s/Mpc with a standard deviation of 0.67 km/s/Mpc, implying a relative error of about 1\% in $H_0$, comparable to the constraining power of Planck and centered  on its \LCDM\ best-fit value. 
We hold the Planck result fixed and do not introduce an extra prior on any of its posteriors. 

We show the result of this exercise in the left panel of Fig.\ \ref{fig: postprocessed} for the $S_{12}\times\omega_m$ subspace constraints as well as the 1-dimensional posterior on $h$, now artificially made tighter for all surveys (usually, their posteriors would be approximately flat within the $h$ plot limits). 
On the top right of both panels, we again compute the $Q_{\rm DM}$ metric as a proxy for the level of agreement between each individual survey and the CMB. 

When the $h$ posterior in the lensing surveys is in agreement with Planck's and similarly constrained, we find that the $Q_{\rm DM}$ metrics indicate a level of tension that closely approaches the reported values in the $S_8\times\Omega_m$ plane (left panel of Fig.\ \ref{fig: baseline}). 
The level of consistency of both DES Y3 and KiDS \textit{vs.}\ CMB reaches a $Q_{\rm DM}\approx 2.5\sigma$, in line with the current paradigm for the lensing amplitude tension as derived from the usual $S_8$ definition. 
This means that, with lensing amplitudes measured over an unambiguous absolute scale of 12 Mpc and $h$ fixed to be in agreement with the CMB value, $S_{12}$ largely recovers the same tension statements that are made with relative distance units as quoted via $S_8$.

No significant change occurs in the $S_8\times\Omega_m$ subspace after importance sampling on $h$ since, as mentioned above, the $S_8$ lensing constraint is by construction largely insensitive to the Hubble parameter.

Next we consider how much tension is obtained if the Hubble parameter on all cosmic shear surveys is externally informed by a prior in \textit{large disagreement} with Planck 2018 and in agreement with a subset of recent local measurements.

\subsection{An External $H_0$ Prior in Agreement with SH0ES}\label{sec: shoes prior}

As a counterpoint, we pick the $H_0$ value measured by the SH0ES collaboration \citep{Riess2021}. 
We again postprocess the MCMC chains of each weak-lensing survey via importance sampling by introducing an informative Gaussian prior on $H_0$ that is now centered at $H_0=73.0$ km/s/Mpc with standard deviation $0.73$ km/s/Mpc, again representing a 1\% relative error that approximates the best fit and constraining power of the local Hubble value of \cite{Riess2021}. 
With this procedure, we quantitatively explore how the weak-lensing posteriors on $S_{12}\times\omega_m$ would shift if the late Universe expands faster than suggested by the CMB.  

The result is shown in the right-hand panel of Fig.\ \ref{fig: postprocessed}, again with $Q_{\rm DM}$ values on the top-right.
We now find a reduction of about $1\sigma$  in the level of tension as seen by the $Q_{\rm DM}$ metric in the $S_{12}\times\omega_m$ subspace, when compared to the reported $S_8$ tension values for each survey (left panel of Fig.\ \ref{fig: baseline}). 
Both DES Y3 and KiDS-1000 now find relative statistical agreement with Planck's constraints in this subspace at the $1.4$ and $1.3\sigma$ levels respectively. 
We also note that in all cases presented, HSC shows overall better agreement with the CMB. 

This exercise clearly introduces a large tension in the full space of cosmological parameters shared by all experiments due to the hidden $h$ dimension,  which we address in Sec. \ref{sec: full parameter space}.  
Additionally, an important caveat unaccounted for in this picture is that Planck's $S_{12}$ posterior should shift if $h$ is different from $0.673$ and the Universe is required to follow $\Lambda$CDM. We shall see in Sec. \ref{sec: LCDM scalings} that a solution to both tensions cannot be obtained at $h=0.73$ in that vanilla model.

In summary, both importance sampling exercises above show that lensing-amplitude-tension statements derived from $S_{12}$ in WL depend on constraints on $h$. 
In what follows, we explore interpretations of this in the context of $\Lambda$CDM and other cosmological models.

\section{Interpretation in $\Lambda$CDM}

\subsection{The Full Parameter Space}\label{sec: full parameter space}

While the tension metrics reported so far have focused  on the 2D subspaces $S_{8}\times\Omega_m$ and $S_{12}\times\omega_m$, it is clear we must account for their dependence on $H_0$. 
Also, in general, tension assessments between experiments must rely on the full space of shared (cosmological) parameters \citep{Lemos2021,Raveri_Doux_2021}, and importantly also benefit from matched priors and modeling choices \citep{Chang2018,Longley2022}. 
Indeed, we can show in what follows that some of the effects seen in Figs.\ \ref{fig: baseline} and \ref{fig: postprocessed} are clarified when looking more broadly at the full cosmological parameter space. 

Focusing on DES as a representative case, we first verify that apart from the subspaces of Figs. \ref{fig: baseline} and \ref{fig: postprocessed} the next parameter direction of interest is $h$. 
The posteriors on other parameters such as the spectral index $n_s$ correlate too weakly with $S_8$ and $S_{12}$ (see discussion in the next Section), and including or excluding those directions from the tension calculations does not make significant differences in $Q_{\mathrm{DM}}$.

While the left- and right-hand panels of Fig.\ \ref{fig: baseline} show very different tension metrics, looking at DES \textit{vs.}\ Planck in the 3D spaces $[S_8, \Omega_m, h]$ and  $[S_{12}, \omega_m, h]$ reveals nearly identical $Q_{\mathrm{DM}}$ values of 2.1$\sigma$. 
Note that this level of tension is relatively close to the $Q_{\mathrm{DM}}=2.4\sigma$ obtained in the 2D space given by $[S_8, \Omega_m]$, but significantly different from the $Q_{\mathrm{DM}}=1.0\sigma$ inferred in the $[S_{12}, \omega_m]$ subspace. 

Additionally, after applying the tight prior $h\approx0.673$ in lensing
as shown in the left-hand panel of Fig.\ \ref{fig: postprocessed}, we find  $Q_{\mathrm{DM}}\approx 2.5\sigma$ in DES \textit{vs.}\ Planck in all of the 3D or 2D subspaces $[S_8, \Omega_m, h]$, $[S_{12}, \omega_m, h]$,   $[S_8, \Omega_m]$ and $[S_{12}, \omega_m]$. 
Finally, with the external prior $h\approx 0.73$ applied to lensing we find, in  shorthand notation: $Q_{\mathrm{DM}}\left([S_8, \Omega_m, h]\right)= Q_{\mathrm{DM}}\left([S_{12}, \omega_m, h]\right)=5.4\sigma$ between DES and Planck, 
while in the 2D spaces of the same experiments there is a clear difference:  $Q_{\mathrm{DM}}\left([S_8, \Omega_m]\right)=2.5\sigma$ \textit{vs.}\ $Q_{\mathrm{DM}}\left([S_{12}, \omega_m]\right)=1.4\sigma$. 
Of course, the larger tension in the 3D space is mainly coming from the discrepancy in $h$ in this case. 
Note again that this does not mean that the main tension in $\Lambda$CDM can be removed by changing $h$ alone,  as Planck's $S_8$ and $S_{12}$ posterior would also change with a shift in $h$. 

The results above reiterate that the correlation between $S_{12}$ and $h$ in the weak-lensing surveys needs to be taken into account for a fair representation of tension between those data sets and the CMB and that $S_{12}$ constraints from lensing surveys are informed by their choice of $h$ prior.  
Below, we explore the reasons for this correlation in $\Lambda$CDM from a semi-analytic perspective.

\subsection{$\Lambda$CDM Scaling Relations}\label{sec: LCDM scalings}

We derive simple scalings between the lensing amplitudes $S_8$ and $S_{12}$  and the Hubble parameter in the context of $\Lambda$CDM to help interpret the results of the previous sections. 

In $\Lambda$CDM,  the dependence of $\sigma_8$ and $S_8$ on different cosmological parameters can be well approximated with a fitting formula presented in \cite{Hu_Jain_2004}.
The functional form of this fit can be understood as follows.
The normalization $A_s$ for the initial curvature power spectrum is set at $k=0.05\,$Mpc$^{-1}$. 
Hence, this must be related to $\sigma_8$ by evolving density perturbations to the present using the matter transfer functions which depend on $\omega_m$, $\omega_b$, and the growth function, and then tilting with $n_s$ to the scale corresponding to $8 h^{-1}\,$Mpc. 

We maintain the functional form, but update the fit coefficients below:
\begin{align}\label{eq:sigma8fit}
    \sigma_{8}\approx & \left(\frac{A_{s}}{3.135\times10^{-9}}\right)^{1/2}\left(\frac{\omega_{b}}{0.024}\right)^{-0.272}\left(\frac{\omega_{m}}{0.14}\right)^{0.513}\nonumber\\
     & \times\left(3.123h\right)^{\left(n_{s}-1\right)/2}\left(\frac{h}{0.72}\right)^{0.698}\left(\frac{G_{0}}{0.76}\right) \,,
\end{align}
with the growth function solution at $z=0$ in $\Lambda$CDM well approximated by:
\begin{equation}\label{eq:G0fit}
    G_0 \approx 0.76\left( \frac{\Omega_m}{0.27}\right)^{0.236}\left(1- 0.014\frac{\sum m_\nu}{0.06\textrm{eV}} \right) \,.
\end{equation}
The updated fitting formula in Eq.~\eqref{eq:sigma8fit} is accurate to  1\% relative error in $\Delta\sigma_8/\sigma_8$ within a 10$\sigma$ range of the Planck 2018 best-fit $\Lambda$CDM parameters ($\omega_b , \omega_m, A_s, h, n_s$) in any direction in this 5D space. 
The fit is improved to within 0.5\% relative error if, instead of the fitting form \eqref{eq:G0fit} for the growth function, the  integral solution in the absence of neutrino mass is used as the baseline:
\begin{align}
    G_{0} =&  \left(\frac{5}{2}\Omega_{m} \int_{0}^{1}\frac{\mathrm{d}a}{\left(aH(a)/H_{0}\right)^{3}}\right) \nonumber\\
    &\times \left(1- 0.014\frac{\sum m_\nu}{0.06\textrm{eV}} \right) \,.
\end{align}

The formulae above can be used to obtain a clear degeneracy direction in the $S_8\times h$ plane for Planck data. 
We consider the $S_8$ definition in Eq.\  \eqref{eq: S8 definition} and approximate $\sigma_8$ using Eq.~\eqref{eq:sigma8fit}. 
The tight measurement of the angular scale $\theta_s$ of the sound horizon in $\Lambda$CDM, indeed the parameter best constrained by the CMB \citep{2001ApJ...549..669H}, fixes $\omega_m h^{1.2}$. 
Reading off the remaining dependence on $h$, this procedure yields approximately:
\begin{equation}\label{eq: Planck S8(h) scaling}
    S_8^{\mathrm{Planck}}(h)\propto h^{-2.3} \quad(\Lambda\mathrm{CDM})
\end{equation}
if $A_s$, $n_s$ and $\omega_b$ are held fixed.

We can similarly obtain a fitting formula for $\sigma_{12}$ in $\Lambda$CDM simply by noting that one must recover $\sigma_8=\sigma_{12}$ at $h=2/3$.
Also valid to  1\% relative error within a 10$\sigma$ range of Planck 2018's best fit, we find 
\begin{align}\label{eq:sigma12fit}
    \sigma_{12}\approx0. & 948\times\left(\frac{A_{s}}{3.135\times10^{-9}}\right)^{1/2}\left(\frac{\omega_{b}}{0.024}\right)^{-0.272}\nonumber\\
    \times & \left(\frac{\omega_{m}}{0.14}\right)^{0.513}\left(2.082\right)^{\left(n_{s}-1\right)/2}\left(\frac{G_{0}}{0.76}\right) \,,
\end{align}
and using the definition \eqref{eq: S12 definition} while fixing $\omega_m h^{1.2}$, $A_s$, $\omega_b$ and $n_s$, we obtain
\begin{equation}\label{eq: Planck S12(h) scaling}
    S_{12}^{\mathrm{Planck}}(h)\propto h^{-1.9} \quad(\Lambda\mathrm{CDM}).
\end{equation}

The scalings presented in equations \eqref{eq: Planck S8(h) scaling} and \eqref{eq: Planck S12(h) scaling} show that $\Lambda$CDM sets a steep dependence between the lensing amplitude parameters and the Hubble value, and that much like $S_8$, the $S_{12}$ parameter also carries the property of ``higher lensing amplitude at lower $H_0$'' and vice-versa to fit the data at fixed $\omega_m h^{1.2}$. 

In an analogous way, we can also approximately predict the scaling between $S_{12}\times h$ for a weak-lensing survey.  
We can write the relationship between $S_{12}$ and $S_8$ implied by the fitting formulae \eqref{eq:sigma8fit} and \eqref{eq:sigma12fit} using the definitions \eqref{eq: S12 definition} and \eqref{eq: S8 definition}:
\begin{equation}\label{eq: S12(h) lensing}
    S_{12} \approx {0.906}\left(\frac{h}{\Omega_m}\right)^{0.1} \left(\frac{3}{2} h\right)^{(1-n_s)/2} S_8. 
\end{equation}
To extract the dependence on $h$, we can identify the best constrained parameter directions in lensing alone. 
In DES particularly, the main constrained direction (derived as a principal component) deviates slightly from the usual $S_8$ definition, and the parameter that is most de-correlated with $\Omega_m$ is $\Sigma^{\mathrm{DES}}_8\equiv \sigma_8 (\Omega_m/0.3)^{0.586}$ (see \citet*{Secco2022} Sec.\ VI.A). 
Considering that fixed parameter, we obtain 
\begin{equation}
    S^{\mathrm{DES}}_{12} \approx {0.817}\frac{h^{0.1}}{\Omega_m^{0.186}} \left(\frac{3}{2} h\right)^{(1-n_s)/2} \Sigma^{\mathrm{DES}}_8. 
\end{equation}
Finally, inspecting the correlation between $\Omega_m \times h$ in the DES cosmic-shear posterior, we fit the direction $\Omega_m  \sim h^{-1}$,  
closely related to the linear power spectrum ``shape parameter'' $\Gamma$ \citep{1992MNRAS.258P...1E,EisensteinHu}.
Reading off the approximate dependence between $S_{12}$ and the Hubble parameter by fixing $\Sigma^{\mathrm{DES}}_8$ and $\Omega_m h$, and taking $n_s \approx 0.96$, we obtain
\begin{equation}\label{eq: DES S12(h) scaling}
    S^{\mathrm{DES}}_{12}\propto h^{0.3} \quad(\Lambda\mathrm{CDM}).
\end{equation}
While Eq.\ \eqref{eq: S12(h) lensing} is general and assumes only $\Lambda$CDM, the constrained directions that led to Eq.\ \eqref{eq: DES S12(h) scaling} are specific to DES and may change with the different coverage of redshift and angular scales utilized by different surveys. 

\begin{figure*}
    \includegraphics[width=\columnwidth]{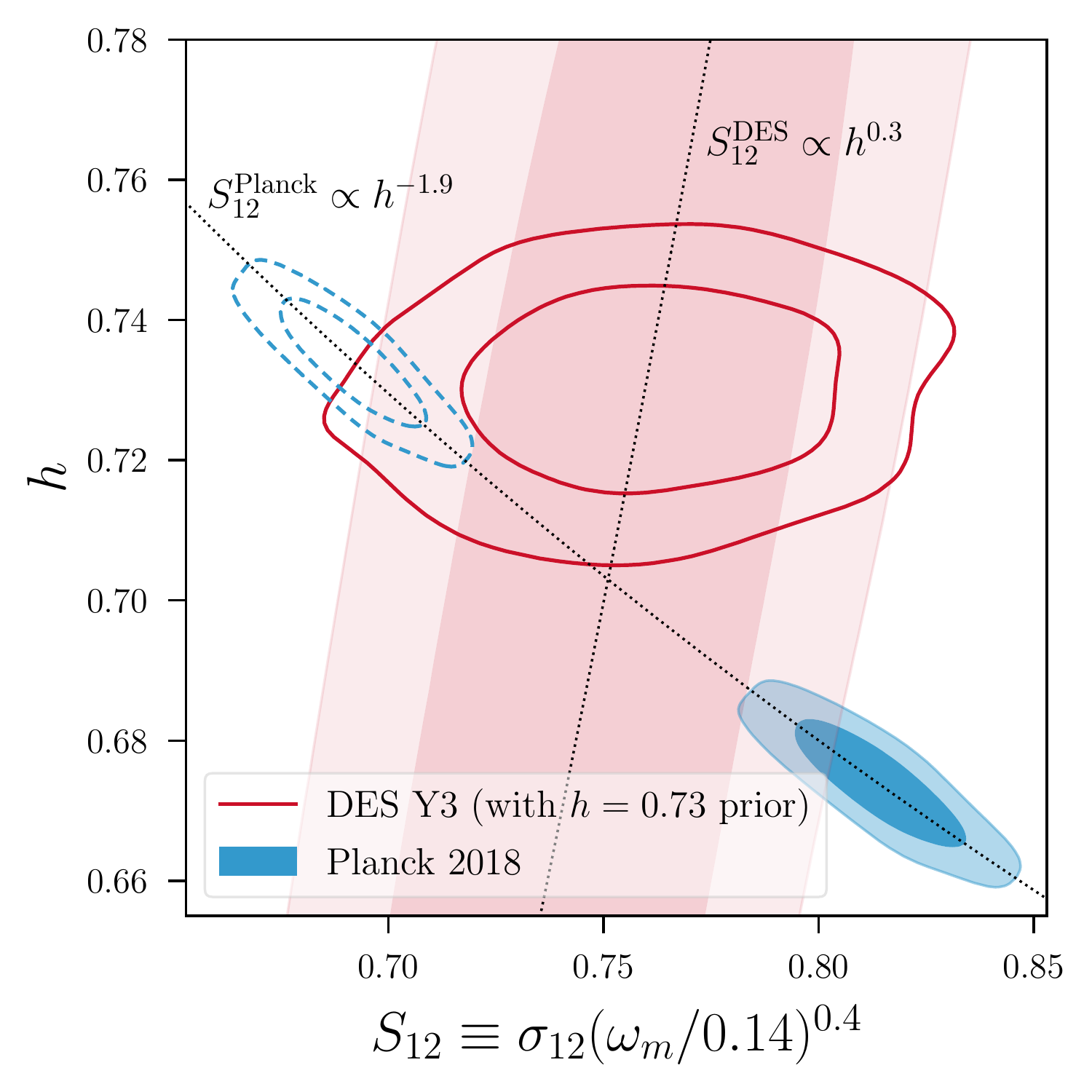}
    \includegraphics[width=\columnwidth]{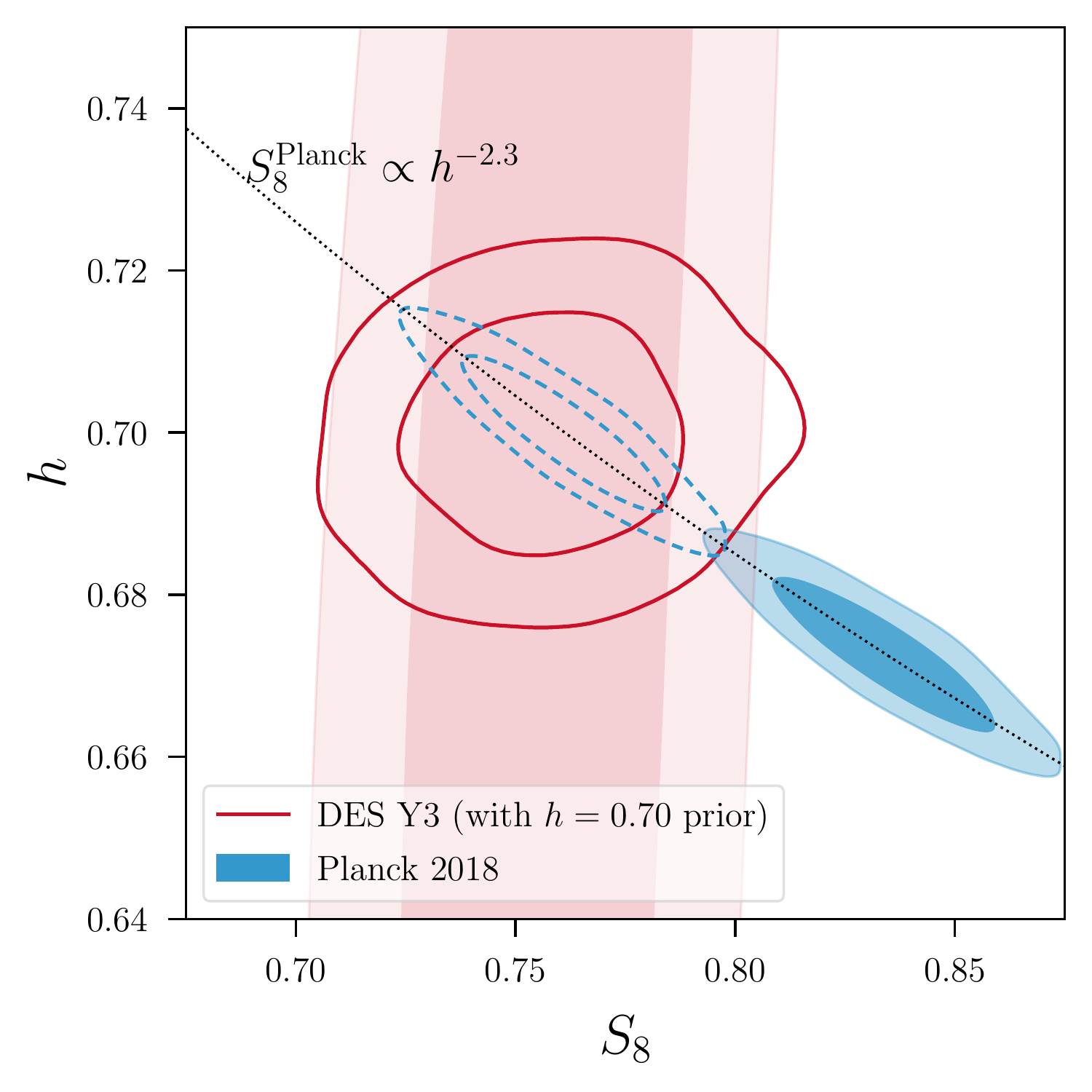}
    \caption{Visualizing the $\Lambda$CDM scaling relations in Eqs.\ \eqref{eq: Planck S8(h) scaling}, \eqref{eq: Planck S12(h) scaling} and \eqref{eq: DES S12(h) scaling}. 
    Red filled contours correspond to DES Y3 without an additional external prior on $h$, solid blue contours show the original Planck \LCDM\ posteriors and dashed blue contours show Planck posteriors that are \textit{artificially shifted} along the $\Lambda$CDM degeneracy lines.  
    \textbf{Left panel:} 
    $S_{12}\times h$ scalings of Planck and DES. The unfilled contours shift the Planck central value to $h=0.73$ and lead to a  smaller $\sim1\sigma$ parameter difference between the Planck and weak lensing, but at high cost of the Planck $\chi^2$ that still reflects the $H_0$ tension.} 
    \textbf{Right panel:} At $h=0.70$, $\Lambda$CDM could in principle accommodate a lower $S_8$ for Planck that is in agreement with weak-lensing surveys (see text for caveats). 
    \label{fig: scalings visualized}
\end{figure*}

The scalings \eqref{eq: Planck S8(h) scaling}, \eqref{eq: Planck S12(h) scaling} and \eqref{eq: DES S12(h) scaling} above can be visualized in Fig.\ \ref{fig: scalings visualized}, which shows Planck and DES posteriors on the $S_{12(8)} \times h$ subspace. 
We note that the $\Lambda$CDM scalings derived semi-analytically closely match the true posterior degeneracy directions.

What this means for the importance-sampling exercises presented in Section \ref{sec: postprocessing} is that a re-parameterization of the lensing amplitude in terms of $S_{12}$ cannot resolve both tensions if $\Lambda$CDM is enforced.
To elaborate, going back to the example of Section \ref{sec: shoes prior} (an $H_0$ prior in agreement with \cite{Riess2021} - right panel of Fig.\ \ref{fig: postprocessed}), 
and artificially ``shifting'' Planck's constraint on $h$ to match the SH0ES result by requiring the plotted posterior to slide along the degeneracy lines of $S_{12}\times h$ in Eq.~\eqref{eq: Planck S12(h) scaling} and $\omega_m \times h$ ($\omega_m\propto h^{-1.2}$) would qualitatively again lead to tension between weak lensing and Planck, with $S^{\mathrm{Planck}}_{12}$ significantly \textit{lower than} lensing surveys. 
The left panel of Fig.\ \ref{fig: scalings visualized} shows this trend by shifting the Planck posterior in the $S_{12}\times h$ plane. 
It is worth emphasizing that these simple shifts along degeneracy directions do not reflect the result of refitting the Planck data with a SH0ES prior, so for that reason we do not quote $Q_{\mathrm{DM}}$ values in this exercise. 

The behavior of the weak-lensing-inferred $S_{12}$ is also easily understood with the scaling in Eqs.\ \eqref{eq: S12(h) lensing} and \eqref{eq: DES S12(h) scaling}. 
Firstly, they suggest a reduction in tension in the $S_{12}\times\omega_m$ subspace at higher $h$, which is indeed what is seen in Fig.\ \ref{fig: postprocessed}. 
Secondly, as referred to in Sec.\ \ref{sec: postprocessing}, the dilution of constraining power in $S_{12}$ comes from its dependence on the poorly-measured $h$ parameter when fixing the best-constrained lensing directions $S_8$ (or  $\Sigma^{\mathrm{DES}}_8$) and $\Omega_m h$. 
That dependence originates both from the shift in physical scale created by the factors of $h$ in Eq.~\eqref{eq: S12(h) lensing}, as well as the shift in amplitude created by the $\Omega_m$ and $\omega_m$ factors in the definitions \eqref{eq: S12 definition} and \eqref{eq: S8 definition} 
at fixed $\Omega_m h$.

More accurate tension measurements could be obtained with actual Monte Carlo sampling of the Planck likelihood with a SH0ES prior, but the qualitative conclusion should be similar: the $\Lambda$CDM scaling cannot solve both the $H_0$ and lensing-amplitude tensions simultaneously if $H_0=73.0$ km/s/Mpc. 
In other words, the resolution of the lensing tension with Planck in \LCDM~would not simply be that Planck is incorrect about $H_0$.

An interesting and related question is: what would be the $H_0$ value for which the vanilla $\Lambda$CDM scalings would bring both lensing and $H_0$ into agreement? 
The answer is a model close to the pre-Planck $\Lambda$CDM concordance \citep{2013ApJS..208...19H}:
the scaling at fixed $\omega_m h^{1.2}$, $\omega_b$, $A_s$ and $n_s$ provided by Eq.~\eqref{eq: Planck S8(h) scaling} brings both DES Y3 and Planck 2018 into qualitative concordance
at around $H_0=70.0$ km/s/Mpc (e.g.\ the result of \cite{Freedman2019} based on a calibration of the tip of the red giant branch). 
The same conclusion would hold for the other Stage-III lensing experiments given the relatively small scatter between them. 
This scenario is shown on the right panel of Figure \ref{fig: scalings visualized}. While this is an interesting perspective, it naturally introduces the puzzle of how could both Planck's and SH0ES's posteriors shift by several standard deviations after the careful treatment of the data and systematics presented by those teams.

The discussion above follows from enforcing \LCDM\ as the underlying cosmological model. Below, we drop that model requirement and explore alternatives qualitatively.

\section{Beyond-$\Lambda$CDM Models}

\subsection{Phenomenological Targets}\label{sec:pheno targets}

We explore  phenomenological scalings of
$S_{8}(h)$ and $S_{12}(h)$ under a new cosmology
that could, in principle, 
shift the Planck lensing amplitude parameters into agreement with WL values. 
We focus still on Planck temperature and polarization only.

Such scalings could be seen as ``theory targets'' 
and their particular forms depend on the value of $H_0$ at which one aims to solve tensions -- as we noted in the previous section for instance, $\Lambda$CDM provides the correct scaling for solving both tensions at $H_0=70.0$ km/s/Mpc.

We create the scaling target in $S_8\times h$ 
simply as a power-law that connects the best-fit Planck $\Lambda$CDM point in that space ($S_8=0.833$, $H_0=67.3$) to the ``concordance'' point given by the best-fit DES Y3 cosmic shear $S_8$ and best-fit SH0ES $H_0$: $(S_8=0.755, H_0=73.0)$. The scaling target for
$S_8$ is then:
\begin{equation}\label{eq: S8 target}
    S^{\mathrm{Planck}}_8(h)\propto h^{-1.2} \quad\mathrm{(non-}\Lambda\mathrm{CDM \,target)}\,. 
\end{equation}
For a less strict agreement to within about a standard deviation from the lensing best fit, the scaling with $h^\beta$ may lie within the range $\beta=[-1.7, -0.7]$, where $\beta<-1.2$ 
underpredicts the Planck $S_8$ value in comparison with lensing, and $\beta>-1.2$ overpredicts it.

Analogously, we define the $S_{12}\times h$ target scaling as a power-law between the Planck best fit in that space ($S_{12}=0.815$, $H_0=67.3$) to the DES+SH0ES point ($S_{12}(h=0.73)=0.753$, $H_0=73.0$), which yields:
\begin{equation}\label{eq: S12 target}
    S^{\mathrm{Planck}}_{12}(h)\propto h^{-1.0} \quad\mathrm{(non-}\Lambda\mathrm{CDM \,target)}\,,
\end{equation}
where we employed the scaling in Eq.\ \eqref{eq: DES S12(h) scaling} to obtain the DES $S_{12}(h=0.73)$ value. The less strict range $\beta=[-1.5, -0.5]$ in $S_{12}(h)\sim h^{\beta}$ provides a qualitative agreement within a standard deviation of DES.

Notice that the scalings  \eqref{eq: S8 target} and \eqref{eq: S12 target}  are different and in particular shallower than $\Lambda$CDM expectation in Eqs.\ \eqref{eq: Planck S8(h) scaling}, \eqref{eq: Planck S12(h) scaling} respectively. It is in this sense that they represent ``non-$\Lambda$CDM targets''. Had we enforced the $\Lambda$CDM scalings and a fit to $H_0=73.0$ km/s/Mpc, the CMB would significantly underpredict the value of $S_8$ in comparison with DES Y3.

Although up to this point we have remained agnostic about the underlying physics of the new cosmological model, it would have to satisfy several important properties due to the caveats in the construction of the scaling relations, which we clarify next.

Firstly, we assume that weak-lensing posteriors do not move under this new cosmology. 
Effectively, this implies that the late universe remains well-described by \LCDM. 
This approach is well-supported by literature that demonstrates that late-universe modifications to background \LCDM\ cosmology cannot resolve the Hubble tension while maintaining a good fit to concordant data \citep{Benevento:2020fev,Efstathiou:2021ocp,Keeley:2022ojz}. 
Hence, this assumption eliminates models that cannot reconcile both tensions simultaneously without introducing new ones.

A related and more specific assumption is that the matter power spectrum should minimally diverge from $\Lambda$CDM 
such that the scaling of $S_{12}^{\rm DES}(h)$ used to derive the target in Eq.\ \eqref{eq: S12 target}, which relies on the shape of the power spectrum via its correlation with $h$, is preserved.

Additionally, recall that the targets are defined as the shift required in the best-fit point of the Planck temperature and polarization data when going from a \LCDM-interpretation to some new cosmology, along a power-law direction.
This encompasses new models that are continuously connected to $\Lambda$CDM by some small parameter, where the relatively small changes in Planck's best fit when extending the model make the bounds in the $S_8\times h$ plane  elongated in the direction of the scalings proposed and provide agreement between data sets.  
More generally, for more radical models, the new, internal $S_8\times h$ degeneracy direction could be different from these scalings which are extrapolated from the baseline $\Lambda$CDM model.   For example, this could occur if what is physically constrained by Planck under the new model implies that $S_8$ is not concordant with its $\Lambda$CDM value $S_8 \approx 0.83$  at its best-fit value for $h \approx 0.67$.

More generally, any beyond-$\Lambda$CDM model that realizes such scalings should also provide a reasonable fit to the CMB data,
for instance via a significant improvement in $\chi^2$ or Bayesian evidence-based comparisons when more degrees-of-freedom are introduced, in comparison with the baseline $\Lambda$CDM best fit. This requirement is far from trivial, and the performance of different candidate models is thoroughly tested in \cite{Schoneberg2022}.

\subsection{EDE as a Test Case}

A model that illustrates the use of these targets and embodies these caveats is ``early dark energy'' (EDE) wherein an extra component increases $H(z)$ only around matter-radiation equality \citep{Karwal2016,Poulin2019,Lin2019,Smith2020,Agrawal:2019lmo}.
It is realized in its simplest form by a scalar field, initially behaving like a cosmological constant and becoming transiently significant to the expansion rate before redshifting away.
This mechanism has been studied as a way to reduce the sound horizon at recombination $r_s$, which has the consequence of increasing the CMB-inferred $H_0$ when the well-measured sound horizon angular scale $\theta_s$ is preserved (\cite{Knox:2019rjx,Aylor:2018drw,Bernal:2016gxb}).

\begin{figure*}
    \includegraphics[width=\columnwidth]{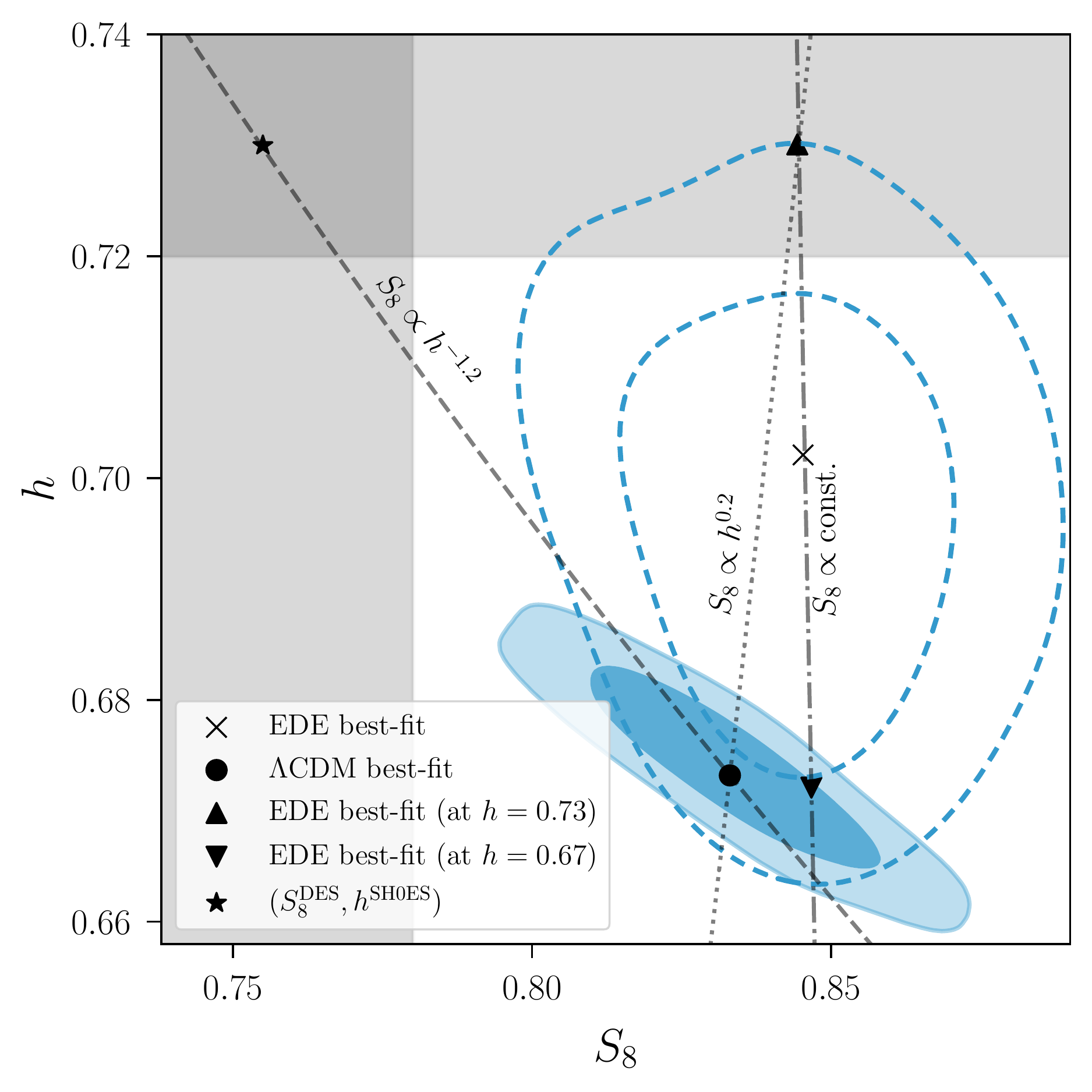}
    \includegraphics[width=\columnwidth]{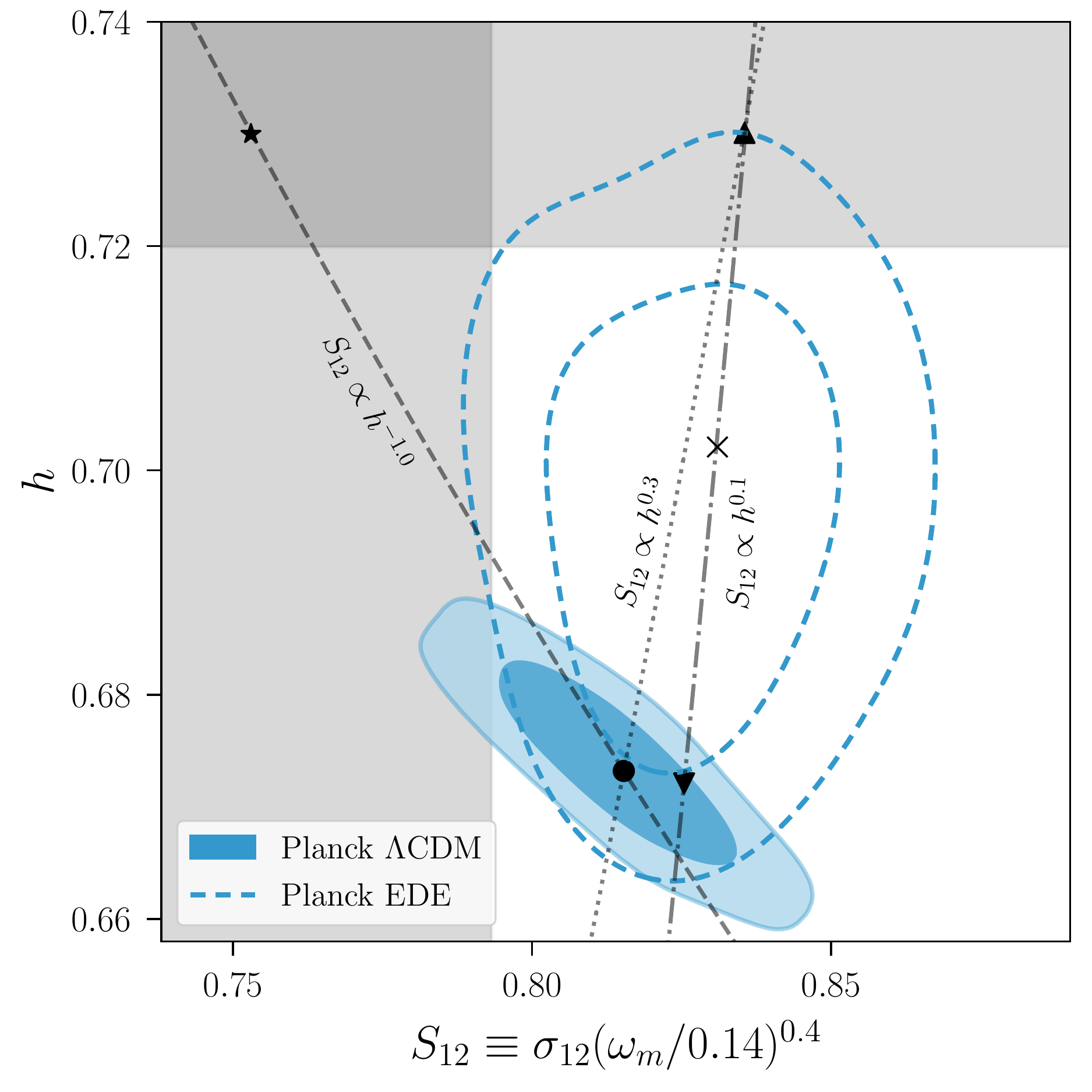}
    \caption{
    Visualizing the EDE and target scaling relations in Eqs.~\eqref{eq: S8 target},~\eqref{eq: S12 target},~\eqref{eq: EDE scaling S8(h)} and~\eqref{eq: EDE scaling S12(h)}, as well as the internal degeneracy of the EDE posterior. 
    In both panels, contours correspond to Planck temperature and polarization $\Lambda$CDM (solid) and EDE (dashed) posteriors. 
    Vertical and horizontal grey bands in both panels correspond to $1\sigma$ uncertainties of DES Y3 cosmic shear's $S_{8(12)}$ and SH0ES's $h$, respectively. 
    \textbf{Left panel:} 
    The target line  $S_8\propto h^{-1.2}$ (dashed) connecting the best-fit values of Planck \LCDM\ at $h=0.67$ 
    and the joint concordance point ($S_{8}^{\mathrm{DES}}\approx0.75$,    $h^{\mathrm{SH0ES}}\approx0.73$), 
    the empirical EDE scaling $S_8\propto h^{0.2}$ (dotted) 
    and the EDE degeneracy direction $S_8\propto \mathrm{constant}$ (dot-dashed), approximately independent of $h$. 
    \textbf{Right panel:} 
    Analogous to the left panel, but showing the target line $S_{12}\propto h^{-1.0}$ (dashed) through the joint concordance point ($S_{12}^{\mathrm{DES}}(h=0.73)\approx0.75$, $h^{\mathrm{SH0ES}}\approx0.73$), the empirical EDE scaling $S_{12}\propto h^{0.3}$ (dotted) and the EDE degeneracy direction $S_{12}\propto h^{0.1}$. 
    }
    \label{fig: EDE and target scalings visualized}
\end{figure*}

Because the EDE component redshifts away soon after equality, the evolution of structure in the late universe in EDE follows that of \LCDM.
The additional information from weak lensing can be mainly characterized through  $S_8$  for the DES Y1 $3\times2$pt data \citep{DESY13x2pt}. 
This is validated in \cite{2020PhRvD.102d3507H}, which finds very small discrepancies in an EDE inference between a full-likelihood treatment of DES Y1 according to \cite{Krause2017} and the use of the DES Y1 \LCDM-based $S_8$ constraint as a Gaussian, independent prior (with small differences owing mostly to $\Omega_m$). 
That is, relative to the constraining power of a combination of other data sets in that work, the DES Y1 $S_8$ posterior for EDE departs minimally from $\Lambda$CDM and remains nearly independent of $H_0$. 
The DES Y3 cosmic shear $S_8$ posterior used as a point of comparison in our targets has an uncertainty comparable to the DES Y1 $3\times2$pt result and is therefore significantly less constraining than Planck 2018 in parameter directions other than the lensing amplitude, so the results of \cite{2020PhRvD.102d3507H} are applicable here.

Furthermore, we check that the EDE low-redshift matter transfer function  shape is nearly preserved when compared to $\Lambda$CDM and the shift induced by $n_s$ is subdominant, which motivates the use of the $S^{\mathrm{DES}}_{12}(h)$ scaling in Eq.\ \eqref{eq: DES S12(h) scaling}. 
In particular, we compare the $S_{12}$ samples in the EDE MCMC result with the right-hand side of Eq.\ \eqref{eq: S12(h) lensing}, computed with the parameters ($S_8$, $\Omega_m$, ...) at the same respective samples. In doing so, we sub-select samples within an approximate $3\sigma$ of the EDE best fit given by a $\Delta \chi^2<9$, simply to highlight the  important samples of 
the parameter space. 

\begin{figure}
    \includegraphics[width=\columnwidth]{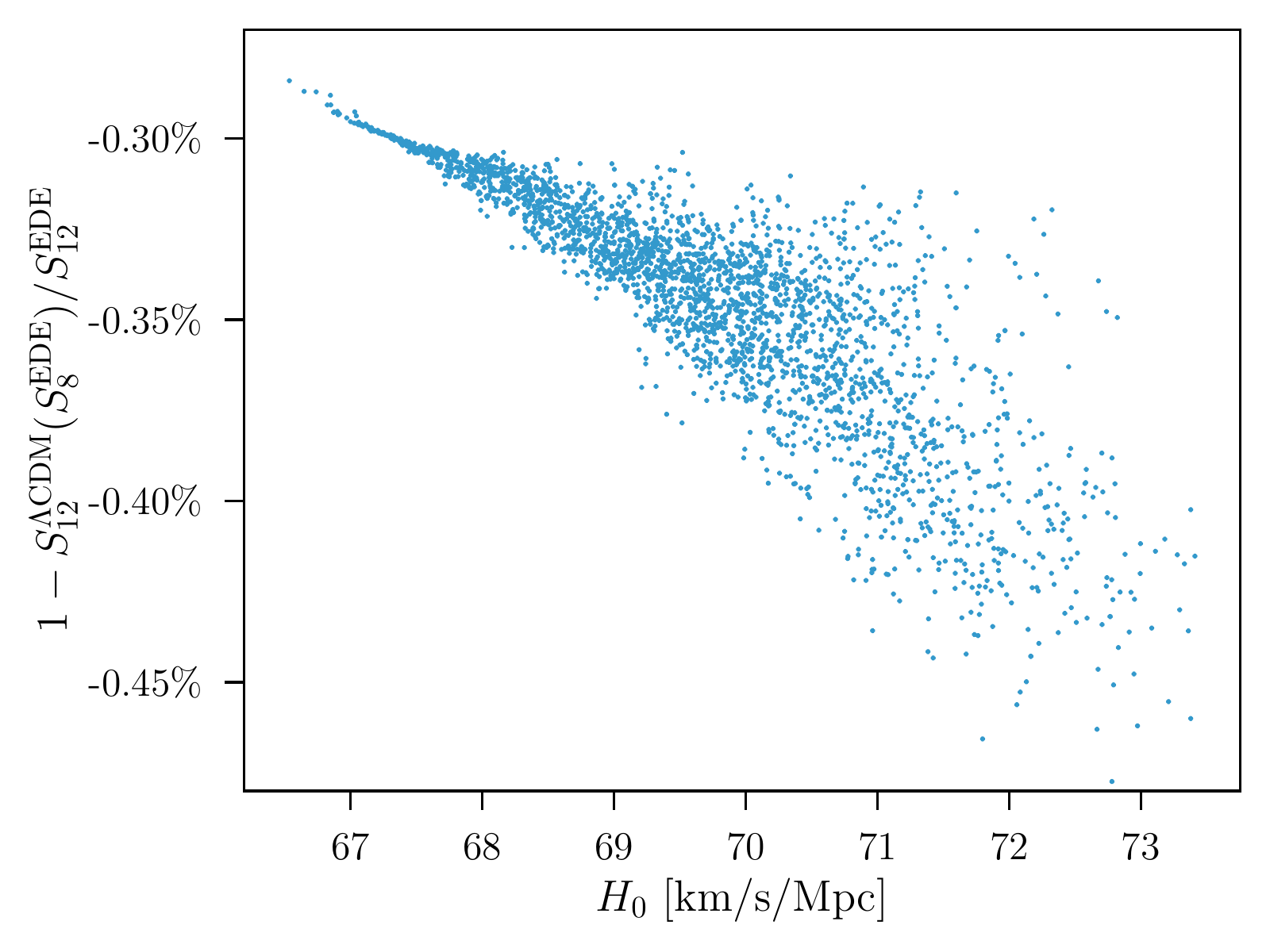}
    \caption{ For an EDE cosmology, we compare 
    the \LCDM\ expectation $S_{12}^{\rm \Lambda CDM}(S^{\mathrm{EDE}}_8)$ utilizing Eq.~\eqref{eq: S12(h) lensing} as a function of the integral result $S^{\mathrm{EDE}}_8$, with the full integral result $S_{12}^{\rm EDE}$. Points are computed at MCMC samples, ordered by their $H_0$ values. Note that the deviations decrease as $H_0$ approaches the $\Lambda$CDM best fit of $H_0\approx 67$ km/s/Mpc but remain less than $0.5\%$ all the way to 73 km/s/Mpc. We thus confirm that the $\Lambda$CDM scaling $S^{\mathrm{DES}}_{12}(h)$ in Eq.~\eqref{eq: DES S12(h) scaling} is sufficiently well-preserved in EDE}
    \label{fig: S12(S8) EDE samples}
\end{figure}

Since Eq.~\eqref{eq: S12(h) lensing} was derived entirely under $\Lambda$CDM, a good agreement in the comparison described above would mean that the relationship between $S_{12}$ and $S_8$  (which we know is preserved from \cite{2020PhRvD.102d3507H}) for a given  EDE cosmology is very close to the $\Lambda$CDM prediction. 
We indeed find such agreement, reported in Fig.\ \ref{fig: S12(S8) EDE samples}, which shows a deviation smaller than $0.5\%$ between the EDE value of $S_{12}$ and its approximation based on the \LCDM\ relation $S_{12}(S_8)$ in Eq.\ \eqref{eq: S12(h) lensing}. 

Next we would like to verify how well EDE obeys the target scalings for Planck  $S_{8}(h)$ and $S_{12}(h)$.   
Note that EDE satisfies the basic requirement for using those scalings in that it is continuously connected to $\Lambda$CDM by a small parameter $f_{\rm ede}$, the maximal fraction of early dark energy density relative to total.  
In order to test the scalings, we sample the posterior distributions of the fundamental EDE parameters 
for Planck 2018 temperature and polarization data (\cite{Planck:2019nip,Planck2018parameters})
with $S_{8}$ and $S_{12}$ as derived parameters. 
We use the AxiCLASS Boltzmann code\footnote{https://github.com/PoulinV/AxiCLASS} for EDE (\cite{Poulin:2018dzj,Smith2020})
and sample using MontePython\footnote{https://github.com/brinckmann/montepython\_public} (\cite{Audren:2012wb, Brinckmann:2018cvx}). 
We follow \cite{Smith:2022hwi} for EDE parameter priors (but with $f_{\rm ede} < 0.03$) and adopt uninformative priors on all \LCDM\ parameters. 

Proceeding with the same definition used for the target scalings, we connect the best-fit LCDM point to the best-fit EDE point at $H_0=73.0$ km/s/Mpc. 
We find this point by optimising the likelihood over all cosmological and nuisance parameters, while holding $H_0$ fixed at its concordant value (equivalent to a profile likelihood approach as e.g.\ \cite{Herold2022}) and find:
\begin{equation}
\label{eq: EDE scaling S8(h)}
    S^{\mathrm{Planck}}_8(h) \propto h^{0.2}
    \quad\mathrm{(EDE)}\,,
\end{equation}
\begin{equation}
\label{eq: EDE scaling S12(h)}
    S^{\mathrm{Planck}}_{12}(h) \propto h^{0.3}
    \quad\mathrm{(EDE)}\,.
\end{equation}

Apparent in the empirical EDE scalings above is that their trend is opposite to the target direction which would solve both tensions jointly, slightly increasing - or at best not alleviating - the statistical significance of the lensing amplitude tension. 
This is a known shortcoming of current implementations of EDE and is attributed primarily to an increase in $\omega_\mathrm{cdm}$ required with EDE to fit CMB data, which in turn increases the lensing amplitude \citep{2020PhRvD.102d3507H, Reeves2022}.

Note that in spite of the trend of higher $S_8$ with higher $h$, the EDE model intrinsically predicts a lower $S_8$ value at a fixed set of cosmological parameters than $\Lambda$CDM as given by Eq.~\eqref{eq:sigma8fit}. 
In EDE, the impact on $S_8$ of the larger $\omega_\mathrm{cdm}$ that accompanies the larger $h$ is offset by the EDE suppression of the early growth of structure.   For example in the maximum likelihood EDE model with $f_{\rm ede}=8.3\%$, $S_8$ is about $3\%$ lower than in $\Lambda$CDM with the same other parameters.

The scalings \eqref{eq: S8 target}-\eqref{eq: EDE scaling S12(h)} and the many best-fit points referenced above can be seen more clearly in Fig.\ \ref{fig: EDE and target scalings visualized}. 
The left panel shows Planck posteriors in $\Lambda$CDM and EDE in the $S_8\times h$ plane along with the directions given by Eqs.\ \eqref{eq: S8 target} and \eqref{eq: EDE scaling S8(h)} and the right panel shows the posteriors in $S_{12}\times h$ along with the target and empirical (EDE) scalings \eqref{eq: S12 target} and \eqref{eq: EDE scaling S12(h)}. 
Horizontal and vertical grey bands show the $1\sigma$ uncertainties of SH0ES and DES Y3 cosmic shear respectively.

Notice also that the internal degeneracy (dot-dashed line) defined by connecting the maximum likelihood EDE models at $h\approx 0.67$ to $h\approx 0.73$ is slightly different than the scaling in Eq.~(\ref{eq: EDE scaling S8(h)}) and in particular $S_8$ is instead nearly independent of $h$. 
This illustrates the caveat that our target scaling connects $\Lambda$CDM at $h\approx 0.67$ to the new model at $h\approx 0.73$ rather than the scaling within the new model.   For EDE this difference of $0.2$ in the exponent is small compared with the deviation from the target scaling of $S_8 \propto h^{-1.2}$.
Similar statements apply to $S_{12}$.

In summary, we have presented above a set of target scalings which are intended to help build models beyond-$\Lambda$CDM 
that  jointly solve the Hubble and lensing-amplitude tensions.  
We have illustrated their use and the assumptions behind them in a model of current interest that solves the Hubble tension, namely EDE. 
The fact that this specific model does not successfully address the lensing-amplitude tension in its current implementation is well known, though quantifying its scalings in comparison with the idealized targets \eqref{eq: EDE scaling S8(h)} and \eqref{eq: EDE scaling S12(h)} may help develop an extension that does. 
We do not find qualitatively different conclusions with the $S_{12}$ parametrization compared to those obtained via $S_8$ -- both parameters seem equally suited to describe fundamental differences between the early- and late-time Universe as seen by the Planck data, though $S_8$ is more convenient for treating lensing alone given its near independence of the constraints on $h$.

\section{Summary \& Conclusions}

Using the latest cosmic shear constraints made public by ongoing weak-lensing surveys (DES Y3, KiDS-1000, HSC-Y1), we have explored a re-definition of the lensing amplitude parameter in terms of the linear matter power spectrum filtered over a 12 Mpc absolute distance scale ($S_{12}$) as proposed by \cite{Sanchez2020}, and compared those survey results against Planck 2018.

When inspecting the 2D plane $S_{12}\times \omega_m$, we find constraints from all lensing surveys to be substantially weakened in this parameterization (Fig.\ \ref{fig: baseline}). 
We trace this loss of constraining power to the marginalization over the Hubble parameter, which is poorly measured in lensing alone and correlates with $S_{12}$ in the absolute distance definition (a feature that is not present in the $S_8$ convention of distances relative to the Hubble length). Simultaneously, we find that the full statistical significance of the lensing amplitude tension measured by $Q_\mathrm{DM}$ in $S_8 \times \Omega_m$ 
moves to the correlated 3D space $S_{12}\times\omega_m\times h$. 
An additional hidden consequence of this correlation is that $S_{12}$ posteriors from lensing surveys depend on their choice of $h$ priors, which are informative despite being wide.

In further exploring the dependence of $S_8$ and $S_{12}$ on the Hubble parameter, we update a fitting formula originally presented in \cite{Hu_Jain_2004} (Eq.\ \ref{eq:sigma8fit}), 
derive its $\sigma_{12}$ analogous form (Eq.\ \ref{eq:sigma12fit}) 
and utilize both to find semi-analytic scalings in $S_8\times h$ and $S_{12}\times h$ of Planck CMB and DES cosmic shear under $\Lambda$CDM (Eqs.\ \ref{eq: Planck S8(h) scaling}, \ref{eq: Planck S12(h) scaling} and \ref{eq: DES S12(h) scaling}). 
Both formulae are good to  1\% or better accuracy within a range of parameters spanning a $10\sigma$ region around the Planck 2018 $\Lambda$CDM best fit. 
These scalings are helpful in explaining the correlations of lensing surveys with $h$ when tighter priors are introduced (Fig.\ \ref{fig: postprocessed}), as well as in showing what joint solutions to the Hubble and lensing tensions would or would not be consistent with $\Lambda$CDM.

Additionally, we obtain idealized target scalings for models aiming to solve both cosmology tensions jointly by dropping the assumption of $\Lambda$CDM, while adjusting only the early-Universe (Eqs.\ \ref{eq: S8 target} and \ref{eq: S12 target}). 
These targets are directions in parameter space which may serve as model-building pointers under a number of caveats mentioned in Sec. \ref{sec:pheno targets}: 
1) low-redshift observations are left unchanged such that no new tensions are introduced, 
2) the matter power spectrum in a new model is similar enough to $\Lambda$CDM that the functional form of $S^{\mathrm{DES}}_{12}$ is approximately preserved, and 
3) the departure from $\Lambda$CDM is ``continuous'' via tuning a parameter (e.g. $f_\textrm{EDE}\to 0$ recovers $\Lambda$CDM).
One example of a model that exhibits this set of properties is early dark energy (EDE), and thus we compare its empirical scalings to the idealized targets,  quantifying this model's known inability in its current formulation to solve the lensing part of the tension (Eqs.\ \ref{eq: EDE scaling S8(h)} and \ref{eq: EDE scaling S12(h)} and Fig.\ \ref{fig: EDE and target scalings visualized}). We find qualitatively similar conclusions with both $S_8$ and $S_{12}$ in this beyond-$\Lambda$CDM setting, and maintain that the approximate independence of $h$ in lensing makes $S_8$ a better-suited parameterization.

Finally, while the absolute-distance convention can provide a coherent picture of the current lensing-amplitude tension, it is complicated in weak lensing alone due to a correlation with the unconstrained Hubble parameter. 
We conclude that $S_8$ remains the most convenient parameterization choice in lensing studies.

\section*{Acknowledgements}

The authors would like to thank
Chihway Chang, 
Colin Hill,
Bhuvnesh Jain,
Evan McDonough,
Vivian Poulin, 
Judit Prat,
Marco Raveri and
Ariel S\'anchez
for useful comments and discussions.
LS was supported by
the Kavli Institute for Cosmological Physics at the University of Chicago through an endowment
from the Kavli Foundation.
W.H.\ 
was supported by U.S.\ Dept.\ of Energy contract DE-FG02-13ER41958 and the Simons Foundation. 
TK was supported by funds provided by the Center for Particle Cosmology, University of Pennsylvania.
EK was supported by the Department of Energy grant DE-SC0020247 and an Alfred P. Sloan Research Fellowship.

\vfill
\bibliographystyle{mnras_2author}
\bibliography{s12bib}

\end{document}